\documentclass{emulateapj}

\usepackage{amsmath}
\usepackage{amssymb}

\newcommand{\Ds}{D_{\mathrm{S}}}
\newcommand{\Dl}{D_{\mathrm{L}}}
\newcommand{\tE}{t_{\mathrm{E}}}

\DeclareMathOperator{\OII}{O-II}
\DeclareMathOperator{\OIII}{O-III}

\slugcomment{Submitted to ApJ \today.}

\shorttitle{Self-lensing models of the Small Magellanic Cloud}
\shortauthors{Mr\'oz \& Poleski}

\begin{document}

\title{New Self-lensing Models of the Small Magellanic Cloud: \\Can Gravitational Microlensing Detect Extragalactic Exoplanets?}

\author{Przemek Mr\'oz$^{1}$\altaffilmark{$\dagger$} and Rados\l{}aw Poleski$^{2}$}
\affil{$^{1}$Warsaw University Observatory, Al. Ujazdowskie 4, 
00-478 Warszawa, Poland}
\affil{$^{2}$Department of Astronomy, Ohio State University, 140 W.
18th Avenue, Columbus, OH 43210, USA}

\altaffiltext{$\dagger$}{Corresponding author: pmroz@astrouw.edu.pl}

\begin{abstract}

We use three-dimensional distributions of classical Cepheids and RR~Lyrae stars in the Small Magellanic Cloud (SMC) to model the stellar density distribution of a young and old stellar population in that galaxy. We use these models to estimate the microlensing self-lensing optical depth to the SMC, which is in excellent agreement with the observations. Our models are consistent with the total stellar mass of the SMC of about $1.0\times 10^9\,M_{\odot}$ under the assumption that all microlensing events toward this galaxy are caused by self-lensing. We also calculate the expected event rates and estimate that future large-scale surveys, like the Large Synoptic Survey Telescope (LSST), will be able to detect up to a few dozen microlensing events in the SMC annually. If the planet frequency in the SMC is similar to that in the Milky Way, a few extragalactic planets can be detected over the course of the LSST survey, provided significant changes in the SMC observing strategy are devised. A relatively small investment of LSST resources can give us a unique probe of the population of extragalactic exoplanets.
\end{abstract}


\section{Introduction}

Since the detection of the first exoplanets \citep{wolszczan,mayor} over 20 years ago, there is growing evidence that exoplanets are ubiquitous in the Milky Way (e.g., \citealt{cassan,howard2012,mayor2011,proxima2016}). However, most of the known extrasolar planets are found in the solar ``neighborhood'' within $\sim 1\,$kpc of the Sun, and current empirical constraints on the planet abundance in different environments are weak (e.g., \citealt{penny2016}). The fundamental cause of this bias toward nearby exoplanets is the fact that most planet-detection techniques rely on the detection of the host light. Since the gravitational microlensing signal does not depend on the brightness of the host, it is one of the techniques that are sensitive to distant planets, both in the Galactic disk and bulge \citep{mao1991,gould1992}. 

The detection of \textit{extragalactic} exoplanets is currently even more challenging. Exploring the ideas of \citet{covono2000} and \citet{baltz2001}, \citet{ingrosso} discussed the possibility of detecting extrasolar planets in M31 using microlensing. Because individual source stars in M31 cannot be resolved by ground-based telescopes, only the brightest sources (i.e., giants) can give rise to detectable microlensing events. \citet{ingrosso} found that, owing to a large angular size of sources, the planetary signal in the majority of events would be ``blurred'' by the finite-source effects. They estimated that planetary deviations can be discovered only in a few percent of detectable events. Pixel lensing surveys of M31 discover only a few events annually \citep{cn2005,dj2006}.
On the other hand, it has been shown that the Large Synoptic Survey Telescope (LSST)-like survey has the potential to discover transits of Jupiter-sized planets in the Magellanic Clouds \citep{lund1,lund2}. It would be nearly impossible, however, to confirm such planets with radial velocities, even with future 30 m class telescopes. 

The Small Magellanic Cloud (SMC) has been the target of microlensing surveys since the 1990s, when -- following the idea proposed by \citet{paczynski} -- three groups (OGLE: \citealt{udalski1993}; MACHO: \citealt{macho1993}; EROS: \citealt{eros1993}) began the search for dark matter in the form of massive compact halo objects (MACHOs).

The most recent results from EROS \citep{tisserand2007}, OGLE-II \citep{wyrzyk2009,wyrzyk2010}, and OGLE-III \citep{wyrzyk2011_L,wyrzyk2011} provide strong upper limits on the MACHO content in the Milky Way halo: less than 9\% of the halo is formed of MACHOs with mass below $1\,M_{\odot}$ (the limit is even stronger for lower masses). On the other hand, current microlensing results cannot exclude MACHOs more massive than several $M_{\odot}$. \citet{wyrzyk2010,wyrzyk2011} measured the microlensing optical depth of $\tau_{\rm O-II}=(1.55 \pm 1.55)\times 10^{-7}$ and $\tau_{\rm O-III}=(1.30 \pm 1.01)\times 10^{-7}$ toward the SMC in the OGLE-II and OGLE-III fields, respectively.

Only five bona fide microlensing events in the SMC have been reported so far, and most (if not all) of them are believed to be due to self-lensing, meaning that both the lens and the source belong to the SMC \citep{sahu1994}. One event, OGLE-2005-SMC-001, can be likely attributed to a halo lens, although the SMC lenses were not definitively ruled out \citep{dong2007}. The microlensing optical depth and event rate in the SMC are larger than in the Large Magellanic Cloud (LMC), because the former is elongated nearly along the line of sight. 

\begin{table}
\caption{Model predictions.}

\begin{tabular}{@{}lc@{}c@{}c@{}c@{}c}
\hline
& $\tau$ & $\langle t_{\rm E,slow}\rangle$ & $\Gamma_{\rm slow}$ & $\langle t_{\rm E,fast}\rangle$ & $\Gamma_{\rm fast}$\\
& ($10^{-7}$) & (day) & ($10^{-7}\,$yr$^{-1}$) & (day) & ($10^{-7}\,$yr$^{-1}$)\\
\hline
OGLE-II & 1.57 & 110 & 3.32 & 79 & 4.62\\
OGLE-III & 1.19 & 113 & 2.45 & 83 & 3.33\\
LSST & 0.83 & 112 & 1.72 & 88 & 2.19\\
all sources & 0.60 & 109 & 1.28 & 92 & 1.52 \\
\hline
\end{tabular}

\textbf{Notes.} Optical depth and event rates scale with the total stellar mass of the SMC $M_*$ as $(M_*/10^9\,M_{\odot})$. Parameters are averaged over sources brighter than $I=21$ (OGLE), $r=24.7$ (LSST), or all sources. \citet{wyrzyk2010} found $\tau_{\OII}=(1.55 \pm 1.55)\times 10^{-7}$ and $\Gamma_{\OII}\approx 5.6\times 10^{-7}\,$yr$^{-1}$ in OGLE-II fields. \citet{wyrzyk2011} measured $\tau_{\OIII}=(1.30 \pm 1.01)\times 10^{-7}$ and $\Gamma_{\OIII}\approx 4.4\times 10^{-7}\,$yr$^{-1}$ in OGLE-III fields.

\label{tab}
\end{table}

Various authors calculated the SMC self-lensing optical depth, usually based on simple analytical approximations of the number density of stars. \citet{palanque1998} and \citet{sahu1998} reported values in the range $(1.0-5.0)\times 10^{-7}$. \citet{graff1999}, using $N$-body simulations of the SMC, found a lower value of $0.4\times 10^{-7}$. In the recent work by \citet{calchinovati2013}, the average optical depth of $0.81\times 10^{-7}$ and $0.39\times 10^{-7}$ was found, toward OGLE-II and OGLE-III fields, respectively.

However, the SMC has an irregular structure, which is difficult to model with analytic approximations (see Section \ref{sec:model}). In this paper, we use three-dimensional distributions of classical Cepheids \citep{ania_cep} and RR Lyrae stars \citep{ania_RR}, based on new observational results from the OGLE survey \citep{soszyn2015,soszyn2016}, to model the density distribution of the SMC and calculate the self-lensing optical depth and event rates (Section \ref{sec:mic}). We also estimate the total stellar mass of the SMC. We show that the LSST-like survey of the SMC is able to discover $20-30$ microlensing events annually, some of which should have planetary anomalies, provided that significant changes of the observing strategy are devised (Section \ref{sec:lsst}). 

\section{Model}
\label{sec:model}

\subsection{Structure}

We assume that the stellar density distribution is the sum of two components: a \textit{young} population, which follows the spatial distribution of classical Cepheids, and an \textit{old} population, tracing the three-dimensional distribution of RR Lyrae stars. Both classical Cepheids and RR Lyrae stars are standard candles and have been recently used to map the structure of the SMC (\citealt{ania_cep,ania_RR,ripepi2017,deb2017,muraveva2017} and references therein). 

Classical Cepheids form a non-planar, extended structure. The galaxy is stretched to nearly 20~kpc and elongated almost along the line of sight \citep{ania_cep, scowcroft2016}. The shape of the young population is not regular and can be best described as an extended ellipsoid with additional substructures.

RR Lyrae stars follow much more regular distribution without any substructures and asymmetries. The spatial distribution of the old population can be approximated as a triaxial ellipsoid with mean axis ratios 1:1.10:2.13, also elongated along the line of sight (\citealt{ania_RR}; see also \citealt{deb2017,muraveva2017}).

We approximate both distributions using a Gaussian mixture model, which is the sum of Gaussian distributions with unknown parameters and weights. The model fitting is performed with the expectation-maximization algorithm \citep{Dempster77}. For both young and old population, we use 32 Gaussian components, which accurately describe the spatial structure of both Cepheids and RR Lyrae stars. We assume, following \citet{bekki2009} and \citet{calchinovati2013}, that the young population comprises 40\% of the total stellar mass of the galaxy.

The microlensing optical depth is directly proportional to the total stellar mass of the SMC, which is not well constrained. Following \citet{bekki2009} and \citet{calchinovati2013}, we assume $M_* = 1.0\times 10^9\,M_{\odot}$ throughout the paper, but one should keep in mind that all derived optical depths scale as $(M_*/10^9\,M_{\odot})$. We find that $M_* = 1.0\times 10^9\,M_{\odot}$ reproduces well the observed microlensing optical depth toward the OGLE fields, providing an independent estimate of the stellar mass of the SMC. \citet{stani2004} report the total stellar mass of $1.8\times 10^9\,M_{\odot}$ (within a radius of 3~kpc), but \citet{vdm2009}, based on the analysis of the same data set, claim a total stellar mass of only $\sim0.31\times 10^9\,M_{\odot}$. 

\subsection{Kinematics}

The kinematics of the SMC is not well measured \citep{vdm2009}. The mean systemic velocity of the SMC is 230~km/s in the east direction and 330~km/s in the south direction \citep{kalli2013} relative to the Sun, based on the multi-epoch data from the \textit{Hubble Space Telescope}. This is consistent with a recent analysis based on \textit{Gaia} data release 1 \citep{vdm2016}. The old and intermediate stellar populations do not show any evidence for systematic rotation. For example, \citet{harris2006} found that velocities of red giant stars are randomly distributed without large velocity gradients (with a dispersion of $27.5\pm 0.5$ km/s). Velocities of young stars \citep{evans2008} show some evidence of systematic rotation (with a gradient of $26.3 \pm 1.6$ km/s/deg). However, the direction of the maximum velocity gradient found by \citet{evans2008} is different from that measured from H\,{\footnotesize I} observations \citep{stani2004}.

\begin{figure*}
\centering
\begin{minipage}[b]{.45\textwidth}
\includegraphics[width=\textwidth]{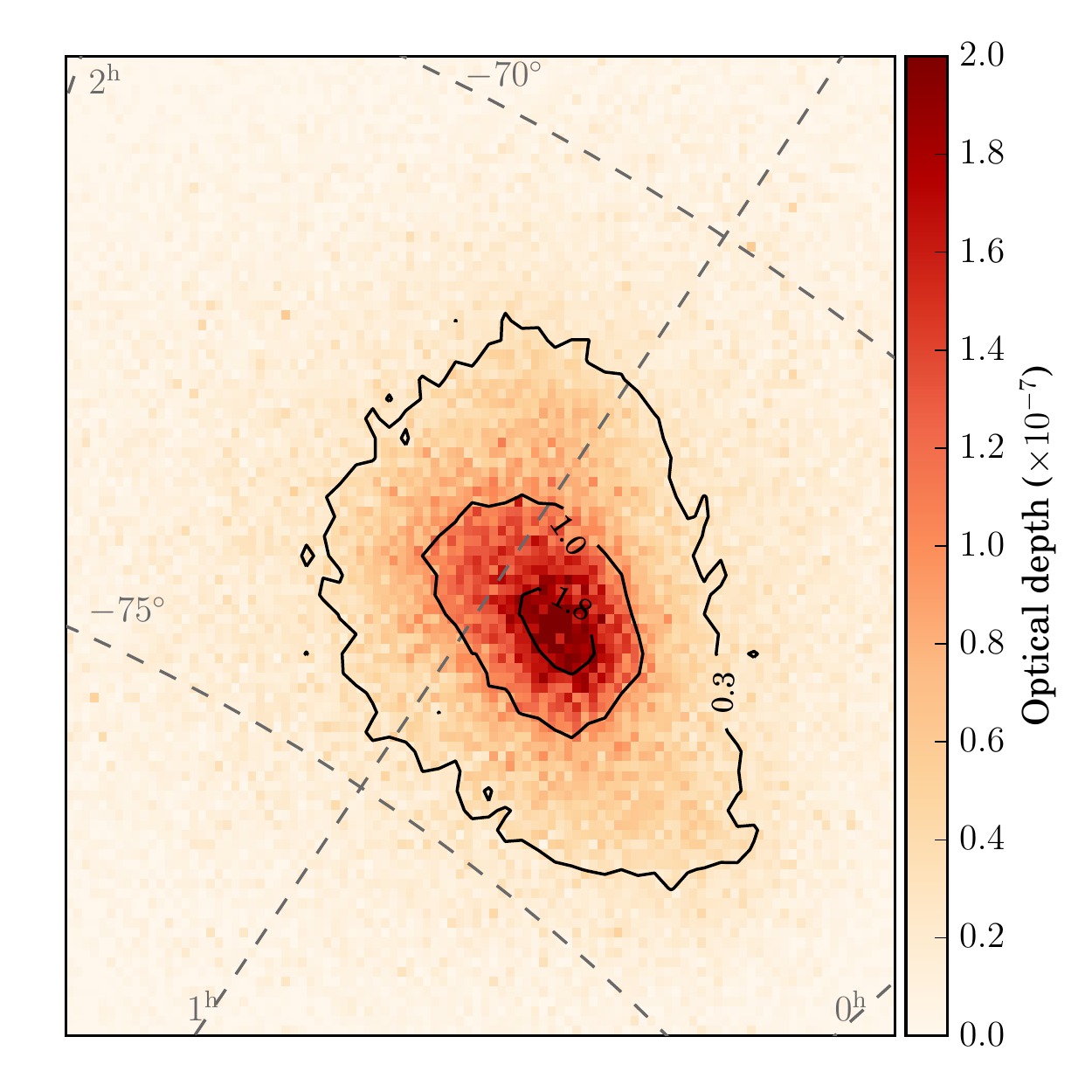}
\caption{Optical depth to self-lensing in the Small Magellanic Cloud, assuming a total stellar mass of $10^{9}\,M_{\odot}$. The map is in the equal-area Hammer projection.\vspace{12pt}}
\label{fig:tau}
\end{minipage}\qquad
\begin{minipage}[b]{.45\textwidth}
\includegraphics[width=\textwidth]{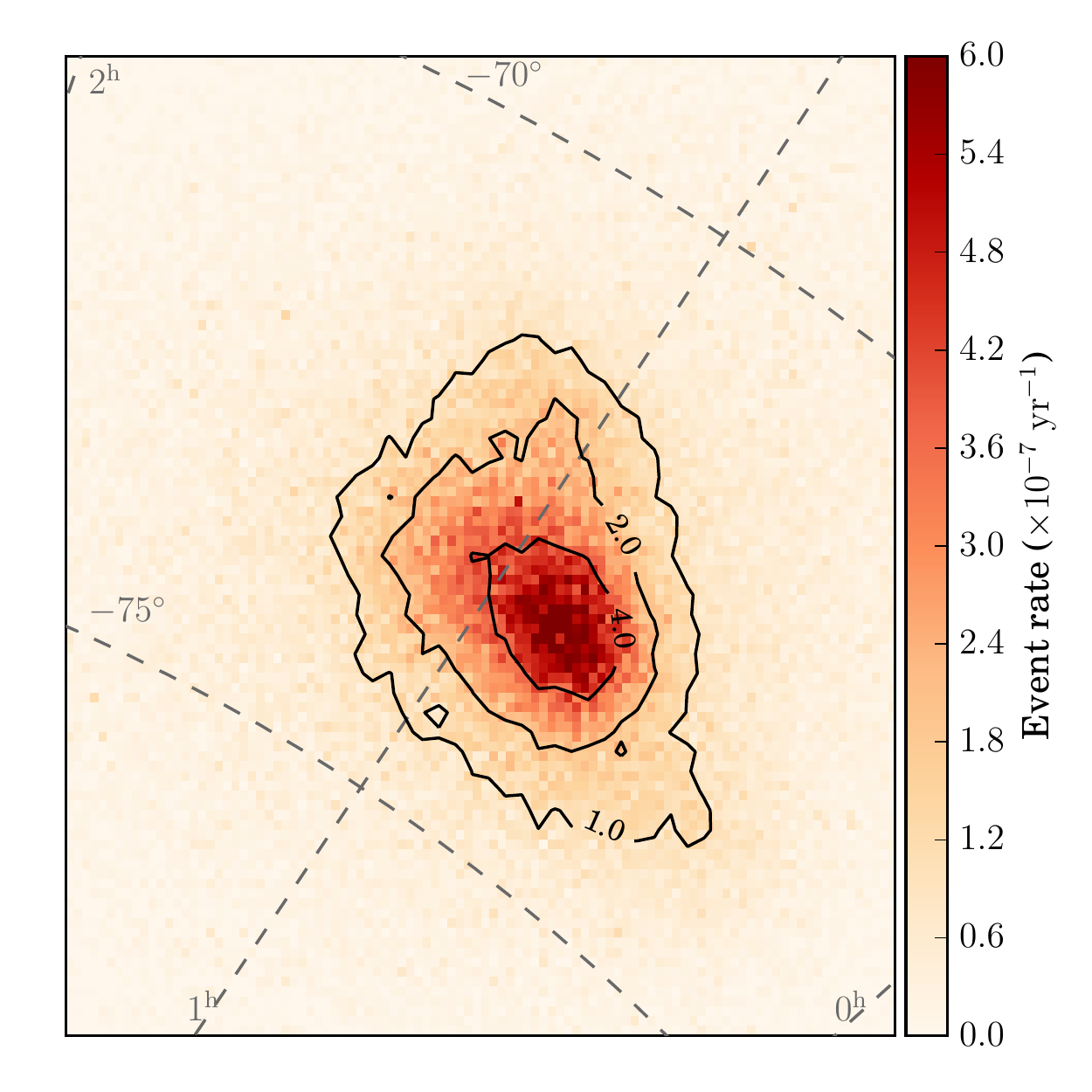}
\caption{Microlensing event rate due to self-lensing in the Small Magellanic Cloud (in the ``fast'' model), assuming a total stellar mass of $10^{9}\,M_{\odot}$. The map is in the equal-area Hammer projection.}
\label{fig:gam}
\end{minipage}
\end{figure*}

For the old population we assume a Gaussian distribution of velocity components with dispersions of 30~km/s in both directions (N and W). Because the kinematics of the young population is not well known, we consider two models. In the ``slow model'' we assign random velocities from the Gaussian distribution with the dispersion of 30~km/s (without rotation). In the ``fast model'' we introduce the rotation around the axis $\boldsymbol{\omega}=(\sin\mathrm{PA},\cos\mathrm{PA},0)$, which is perpendicular to the direction of the maximum radial velocity gradient found by \citet{evans2008} and to the line-of-sight direction. The Cartesian coordinate system has the origin in the center of the mass of young population and its versors are defined as in \citet{ania_cep}. Here, $\mathrm{PA}=126^{\circ}$ is the position angle of the maximum velocity gradient line from \citet{evans2008}. We assume that the rotation velocity is increasing linearly up to 60 km/s with a turnover radius at 3~kpc and we add random components with the dispersion of 30 km/s \citep{stani2004,evans2008}.

\subsection{Initial Mass Function and Isochrones}
\label{sec:imf}

We assume a \citet{kroupa2001} initial mass function with slopes $0.3$ for brown dwarfs ($0.01<M/M_{\odot}<0.08$), $1.3$ for low-mass stars ($0.08<M/M_{\odot}<0.5$), and $2.3$ for high-mass stars ($0.5<M/M_{\odot}<150$). Brown dwarfs constitute 37\% of all objects (about 5\% of the total stellar mass). The mean mass is $0.38\ M_{\odot}$, and so the number of stars in our fiducial model is $0.63M_* / 0.38 = 1.66\times 10^{9}$. Masses of stellar remnants are calculated following the approach of \citet{gould2000} and prescriptions of \citet{mroz2017}.

To estimate the number of sources that could be monitored by a given survey, we generated theoretical isochrones for 200~Myr (young population) and 10~Gyr (old population) for the SMC metallicity ($Z=0.004$) using PARSEC-COLIBRI models \citep{marigo2017}. We assume the foreground reddening of $E(V-I)=0.04$ \citep{haschke2011}.

\section{Microlensing Predictions}
\label{sec:mic}

\subsection{Optical Depth}

The microlensing optical depth toward a given source located at a distance $D_{\rm S}$ is \citep{kiraga1994}
\begin{equation}
\tau(\Ds) = \frac{4\pi G}{c^2}\int_0^{\Ds}\rho_{\rm L}(\Dl) \frac{\Dl(\Ds-\Dl)}{\Ds}d\Dl.
\end{equation}
Here, $\Dl$ is the distance to the lens and $\rho_{\rm L}$ is the density of lenses. In theory, the optical depth depends only on the mass distribution and is independent of other model assumptions (mass function, kinematics). As we show below, however, the optical depth depends also on the limiting magnitude of the survey, and thus on the star-formation history and the mass function of the SMC. We evaluate the mean optical depth $\langle\tau\rangle$ by drawing random sources from our fiducial model and averaging $\tau$ over all sources. The resulting optical depth map in the equal-area Hammer projection is shown in Fig. \ref{fig:tau}.

We do not take into account the microlensing by the Galactic halo objects and we neglect the contribution of nearby Galactic disk lenses. As shown by \citet{calchinovati2013}, the expected signal from disk lenses is $10-20$ times smaller than the SMC self-lensing signal.

The optical depth in our model (Table \ref{tab}) is in excellent agreement with OGLE-II and OGLE-III results \citep{wyrzyk2010,wyrzyk2011}: $\tau_{\OII}=(1.55 \pm 1.55)\times 10^{-7}$ (based in one event) and $\tau_{\OIII}=(1.30 \pm 1.01)\times 10^{-7}$ (based on three events). We would like to point out that \citet{wyrzyk2010,wyrzyk2011} measured the optical depth averaged over sources brighter than $I=21$, which constitute 2.7\% of young stars and only 0.2\% of old stars. The optical depth in our model, averaged over sources brighter than $I=21$, is $\tau_{\OII}=1.57\times 10^{-7}$ and $\tau_{\OIII}=1.19 \times 10^{-7}$ for the total stellar mass of $M_* = 1.0\times 10^9\,M_{\odot}$, in excellent agreement with observations. The optical depth in the same fields, but averaged over all sources, is $0.76-0.87$ times smaller. However, models are not well constrained by empirical estimates, which are based on a small statistics, and there are some arguments that at least one of the SMC events is not due to self-lensing \citep{dong2007,calchinovati2013}. The analysis of new observations from the OGLE-IV survey \citep{udalski2015} should provide us with stronger constraints. The optical depth to self-lensing in OGLE fields in our models is a factor of $2-3$ larger than the numbers calculated by \citet{calchinovati2013} for the same stellar mass of the SMC, likely because we averaged $\tau$ over ``bright'' sources and the observed line-of-sight length of the SMC is larger than that adopted in analytical models by \citet{calchinovati2013}.

\newpage
\subsection{Event Rate}

The differential event rate toward a given source is \citep{batista2011,clanton2014}
\begin{equation}
\frac{d^4\Gamma}{dD_{\rm L}dM_{\rm L}d^2\boldsymbol{\mu}} = 2R_{\rm E}v_{\rm rel}n(D_{\rm L}) f(\boldsymbol{\mu}) g(M_{\rm L}),
\label{eq:gamma}
\end{equation}
where $R_{\rm E}=\sqrt{\frac{4GM_{\rm L}}{c^2}\frac{(D_{\rm S}-D_{\rm L})D_{\rm L}}{D_{\rm S}}}$ is the physical Einstein radius, $M_{\rm L}$ is the lens mass, $n(D_{\rm L})$ is the local number density of lenses, $v_{\rm rel}=|\boldsymbol{\mu}|D_{\rm L}$ is the lens-source relative velocity, $f(\boldsymbol{\mu})$ is the two-dimensional probability density for a given lens-source relative proper motion $\boldsymbol{\mu}$, and $g(M_{\rm L})$ is the mass function. The total event rate can be obtained by integrating Eq. (\ref{eq:gamma}) and averaging over all sources. Eq. (\ref{eq:gamma}) can also be used for generating a random ensemble of microlensing events from our model. The procedure is described meticulously by \citet{clanton2014}. In short: (1) We draw a random source located at distance $\Ds$ and equatorial coordinates $(\alpha,\delta)$ from our fiducial distribution. (2) We draw a random lens at distance $\Dl$ from the range $[0,\Ds]$ from the density distribution in the given direction $n(D_{\rm L},\alpha,\delta)$. (3) We assign random velocities of the lens and the source from Gaussian distributions and calculate the relative velocity $v_{\rm rel}$. (4) We draw a random lens mass $M_{\rm L}$ from the mass function (taking into account stellar remnants). Mass functions of young and old stellar populations are slightly different, but this effect has negligible impact on the calculated timescales. (5) We evaluate the Einstein radius $R_{\rm E}$ and the event timescale $t_{\rm E}=R_{\rm E}/v_{\rm rel}$. For each event we assign a weight $w=R_{\rm E}v_{\rm rel}$. We then estimate the mean timescale from $\langle\tE\rangle = \sum w_i t_{{\mathrm E},i} / \sum w_i$. The mean event rate is simply
\begin{equation}
\Gamma = \frac{2}{\pi}\frac{\tau}{\langle\tE\rangle}.
\end{equation}
Contrary to the optical depth, the event rate depends on the mass function of lenses and their kinematics. Fig. \ref{fig:gam} shows the predicted map of the event rate in the ``fast model.''

\begin{figure}
\centering
\includegraphics[width=0.5\textwidth]{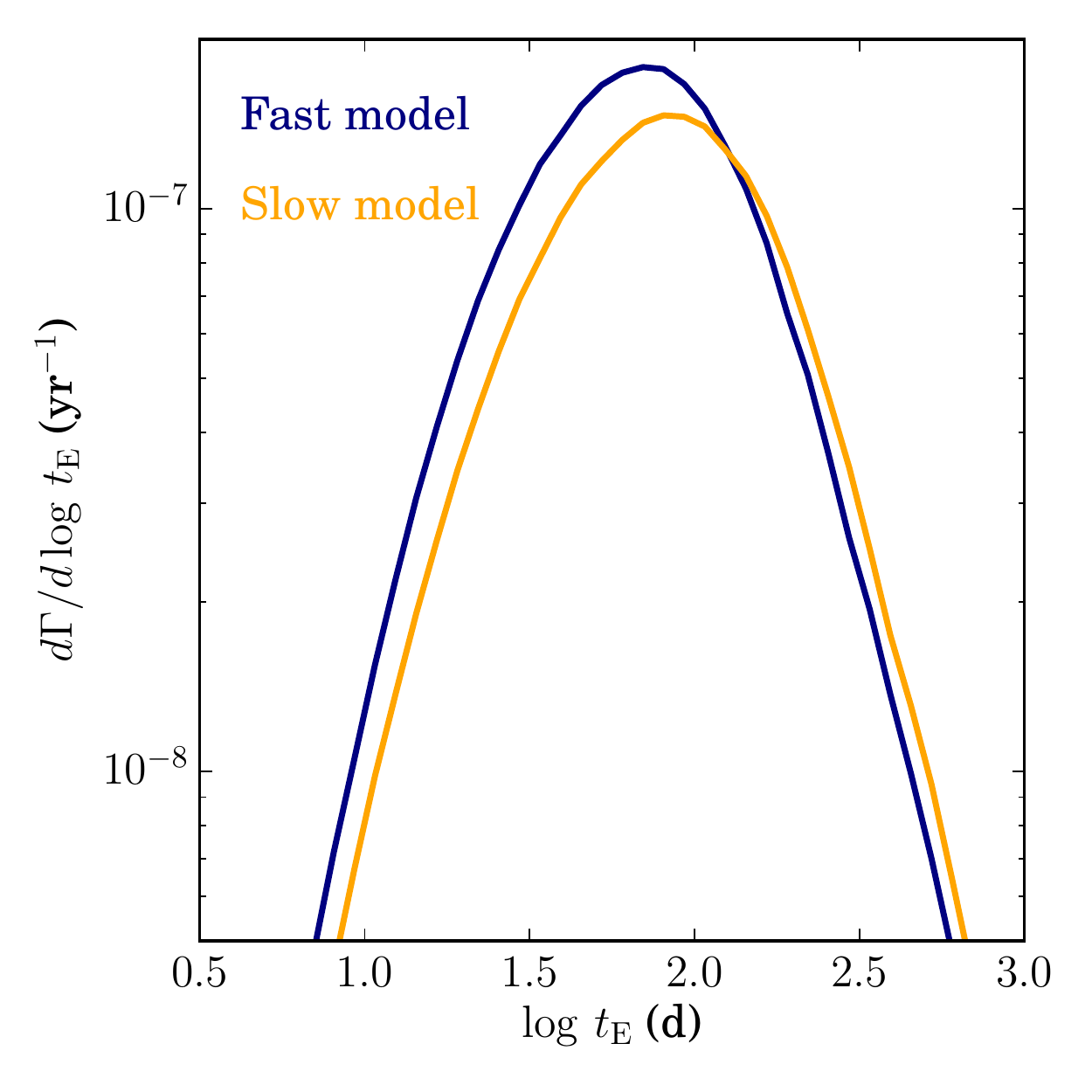}
\caption{The distribution of event timescales in fast and slow models (for the total stellar mass of $10^{9}\,M_{\odot}$).}
\label{fig:tE}
\end{figure}

The mean event timescale in our models ($\langle\tE\rangle=109\,$days in the slow model and $\langle\tE\rangle=92\,$days in the fast model) is similar to that found by \citet{calchinovati2013}, although slightly longer than the mean timescales of events ($67\pm36\,$days) found by \citet{wyrzyk2010,wyrzyk2011}. Fig. \ref{fig:tE} shows the distribution of event timescales in fast and slow models. On average, timescales are $\sim 3$ times longer than those observed toward the Galactic bulge. The mean angular Einstein radius in the ``fast'' model is equal to 60~$\mu$as (with a 68\% confidence interval of $24-96\,\mu$as). Since the typical angular radius of the solar-like main-sequence sources in the SMC is $\sim 0.07\,\mu$as, the finite-source effect does not reduce the planet sensitivity \citep{ingrosso}.

\section{Can Gravitational Microlensing Detect Exoplanets in the SMC?}
\label{sec:lsst}

The upcoming Large Synoptic Survey Telescope (LSST), as we show below, has the potential to observe $20-30$ microlensing events in the SMC annually. However, the microlensing experiment would require significant changes to the default observing strategy to detect and characterize exoplanets in the SMC. The telescope will have a field of view of 9.6~deg$^2$ (which is sufficient to cover the central regions of the SMC with only a single LSST field) and will be able to detect objects as faint as $r=24.7$ (AB magnitudes) in a single-visit image (see LSST Science Book; \citealt{LSST}). The survey is expected to last ten years.

To estimate the number of sources that could be monitored by the LSST, we generated theoretical isochrones (see Section \ref{sec:imf}). The LSST is expected to register sources brighter than $r=24.7$ (absolute magnitude of 5.8), i.e., in the mass range $[0.72,0.93]\,M_{\odot}$ (old population) and $[0.78,3.76]\,M_{\odot}$ (young population). Such stars comprise 7.2\% of all stars in the SMC. This fraction is slightly smaller in $g$ and $i$ bands (given limiting magnitudes of $g=25.0$ and $i=24.0$; \citealt{LSST}). The expected observed event rate is simply the product of the mean event rate $\Gamma$ and the number of observable sources. It ranges from 20 to 26 per year,  for ``slow'' and ``fast'' models, respectively.

The average number of epochs ($\sim 1000$ over 10 years in all filters) in LSST fields is not sufficient to detect ongoing microlensing events and planetary anomalies. However, some fields are planned to be observed with a much higher cadence. We propose that one such high-cadence field should cover the central regions of the SMC. We also propose to observe this field, ideally, in single filter ($r$) each night, which is sufficient to trigger dense follow-up observations. However, such strategy requires an efficient system of real-time detections \citep{udalskiews} and rapid-response follow-up telescopes. High-magnification events, which are very sensitive to planets \citep{griest1998}, can be followed up with smaller telescopes. The number of magnified events at any moment of time is small enough to warrant relatively small telescope-time demand. High-cadence observations also make it possible to remove virtually all false positives, like dwarf novae or background supernovae. 

Another possible strategy is to observe the SMC field every $1-2$ hr, which is sufficient to cover short-timescale planetary anomalies. Such an approach is adopted by current microlensing surveys of the Galactic bulge and allows discovering and characterizing planetary anomalies without the need of targeted follow-up observations. If neither of the above proposed strategies is adopted, a significantly smaller number of SMC events would be discovered. We estimate that over 80\% of the SMC events could be detected in real-time (meaning that at least three consecutive data points deviate $3\sigma$ above the baseline before an event peaks) for a 1 day survey cadence. This number drops dramatically for lower cadence, for example, less than 30\% of events could be detected in real-time if observations are conducted every 20 days.

How many planetary events are expected? \citet{yossi2016} conducted a ``second generation'' microlensing survey for extrasolar planets toward the Galactic bulge based on observations from OGLE, MOA, and Wise. They found that over 12\% of analyzed events showed deviations from single-lens microlensing, and in about one-third of those the anomaly was likely caused by a planet. Most probably, LSST observations of the SMC will not be able to achieve a 4\% success rate, unless high-cadence follow-up is conducted. \citet{yossi2016} used nearly continuous observations of the Galactic bulge, while the SMC can be observed at most $\sim 7$~hr each night (above airmass 1.6). On the other hand, microlensing events in the SMC are on average three times longer than events observed toward the Galactic bulge, making the planetary deviations also longer. 

If we assume that 1--2\% of SMC events have planetary anomalies (and take into account that the SMC can be observed during $\sim2/3$ of a year), then the expected rate of planet detections is $0.1-0.5\,$yr$^{-1}$. A few extragalactic planets should be detected over the course of the LSST (or similar) survey.
Besides detecting the microlensing events, high-cadence observations of the SMC can provide the opportunity to study the stellar variability on all timescales, stellar populations, star-formation and chemical abundance history of the SMC \citep{szkody}, or detect planetary transit candidates \citep{lund1,lund2}. We note that the SMC is better target for microlensing survey than the LMC because the former is elongated along the line of sight. The two galaxies do not differ that much as transit survey targets.

\section*{Acknowledgements}

We thank A. Udalski, S. Koz\l{}owski, and J. Skowron for reading the manuscript.

\bibliographystyle{aasjournal}
\bibliography{pap}

\begin{thebibliography}{}
\expandafter\ifx\csname natexlab\endcsname\relax\def\natexlab#1{#1}\fi

\bibitem[{{Alcock} {et~al.}(1993){Alcock}, {Akerlof}, {Allsman}, {Axelrod},
  {Bennett}, {Chan}, {Cook}, {Freeman}, {Griest}, {Marshall}, {Park},
  {Perlmutter}, {Peterson}, {Pratt}, {Quinn}, {Rodgers}, {Stubbs}, \&
  {Sutherland}}]{macho1993}
{Alcock}, C., {Akerlof}, C.~W., {Allsman}, R.~A., {et~al.} 1993, \nat, 365, 621

\bibitem[{{Anglada-Escud{\'e}} {et~al.}(2016){Anglada-Escud{\'e}}, {Amado},
  {Barnes}, {Berdi{\~n}as}, {Butler}, {Coleman}, {de La Cueva}, {Dreizler},
  {Endl}, {Giesers}, {Jeffers}, {Jenkins}, {Jones}, {Kiraga}, {K{\"u}rster},
  {L{\'o}pez-Gonz{\'a}lez}, {Marvin}, {Morales}, {Morin}, {Nelson}, {Ortiz},
  {Ofir}, {Paardekooper}, {Reiners}, {Rodr{\'{\i}}guez},
  {Rodr{\'{\i}}guez-L{\'o}pez}, {Sarmiento}, {Strachan}, {Tsapras}, {Tuomi}, \&
  {Zechmeister}}]{proxima2016}
{Anglada-Escud{\'e}}, G., {Amado}, P.~J., {Barnes}, J., {et~al.} 2016, \nat,
  536, 437

\bibitem[{{Aubourg} {et~al.}(1993){Aubourg}, {Bareyre}, {Br{\'e}hin}, {Gros},
  {Lachi{\`e}ze-Rey}, {Laurent}, {Lesquoy}, {Magneville}, {Milsztajn},
  {Moscoso}, {Queinnec}, {Rich}, {Spiro}, {Vigroux}, {Zylberajch}, {Ansari},
  {Cavalier}, {Moniez}, {Beaulieu}, {Ferlet}, {Grison}, {Vidal-Madjar},
  {Guibert}, {Moreau}, {Tajahmady}, {Maurice}, {Pr{\'e}v{\^o}t}, \&
  {Gry}}]{eros1993}
{Aubourg}, E., {Bareyre}, P., {Br{\'e}hin}, S., {et~al.} 1993, \nat, 365, 623

\bibitem[{{Baltz} \& {Gondolo}(2001)}]{baltz2001}
{Baltz}, E.~A., \& {Gondolo}, P. 2001, \apj, 559, 41

\bibitem[{{Batista} {et~al.}(2011){Batista}, {Gould}, {Dieters}, {Dong},
  {Bond}, {Beaulieu}, {Maoz}, {Monard}, {Christie}, {McCormick}, {Albrow},
  {Horne}, {Tsapras}, {Burgdorf}, {Calchi Novati}, {Skottfelt}, {Caldwell},
  {Koz{\l}owski}, {Kubas}, {Gaudi}, {Han}, {Bennett}, {An}, {MOA
  Collaboration}, {Abe}, {Botzler}, {Douchin}, {Freeman}, {Fukui}, {Furusawa},
  {Hearnshaw}, {Hosaka}, {Itow}, {Kamiya}, {Kilmartin}, {Korpela}, {Lin},
  {Ling}, {Makita}, {Masuda}, {Matsubara}, {Miyake}, {Muraki}, {Nagaya},
  {Nishimoto}, {Ohnishi}, {Okumura}, {Perrott}, {Rattenbury}, {Saito},
  {Sullivan}, {Sumi}, {Sweatman}, {Tristram}, {von Seggern}, {Yock}, {PLANET
  Collaboration}, {Brillant}, {Calitz}, {Cassan}, {Cole}, {Cook}, {Coutures},
  {Dominis Prester}, {Donatowicz}, {Greenhill}, {Hoffman}, {Jablonski}, {Kane},
  {Kains}, {Marquette}, {Martin}, {Martioli}, {Meintjes}, {Menzies},
  {Pedretti}, {Pollard}, {Sahu}, {Vinter}, {Wambsganss}, {Watson}, {Williams},
  {Zub}, {FUN Collaboration}, {Allen}, {Bolt}, {Bos}, {DePoy}, {Drummond},
  {Eastman}, {Gal-Yam}, {Gorbikov}, {Higgins}, {Janczak}, {Kaspi}, {Lee},
  {Mallia}, {Maury}, {Monard}, {Moorhouse}, {Morgan}, {Natusch}, {Ofek},
  {Park}, {Pogge}, {Polishook}, {Santallo}, {Shporer}, {Spector}, {Thornley},
  {Yee}, {MiNDSTEp Consortium}, {Bozza}, {Browne}, {Dominik}, {Dreizler},
  {Finet}, {Glitrup}, {Grundahl}, {Harps{\o}e}, {Hessman}, {Hinse},
  {Hundertmark}, {J{\o}rgensen}, {Liebig}, {Maier}, {Mancini}, {Mathiasen},
  {Rahvar}, {Ricci}, {Scarpetta}, {Southworth}, {Surdej}, {Zimmer}, {RoboNet
  Collaboration}, {Allan}, {Bramich}, {Snodgrass}, {Steele}, \&
  {Street}}]{batista2011}
{Batista}, V., {Gould}, A., {Dieters}, S., {et~al.} 2011, \aap, 529, A102

\bibitem[{{Bekki} \& {Chiba}(2009)}]{bekki2009}
{Bekki}, K., \& {Chiba}, M. 2009, \pasa, 26, 48

\bibitem[{{Calchi Novati} {et~al.}(2013){Calchi Novati}, {Mirzoyan}, {Jetzer},
  \& {Scarpetta}}]{calchinovati2013}
{Calchi Novati}, S., {Mirzoyan}, S., {Jetzer}, P., \& {Scarpetta}, G. 2013,
  \mnras, 435, 1582

\bibitem[{{Calchi Novati} {et~al.}(2005){Calchi Novati}, {Paulin-Henriksson},
  {An}, {Baillon}, {Belokurov}, {Carr}, {Cr{\'e}z{\'e}}, {Evans},
  {Giraud-H{\'e}raud}, {Gould}, {Hewett}, {Jetzer}, {Kaplan}, {Kerins},
  {Smartt}, {Stalin}, {Tsapras}, {Weston}, \& {Point-Agape
  Collaboration}}]{cn2005}
{Calchi Novati}, S., {Paulin-Henriksson}, S., {An}, J., {et~al.} 2005, \aap,
  443, 911

\bibitem[{{Cassan} {et~al.}(2012){Cassan}, {Kubas}, {Beaulieu}, {Dominik},
  {Horne}, {Greenhill}, {Wambsganss}, {Menzies}, {Williams}, {J{\o}rgensen},
  {Udalski}, {Bennett}, {Albrow}, {Batista}, {Brillant}, {Caldwell}, {Cole},
  {Coutures}, {Cook}, {Dieters}, {Dominis Prester}, {Donatowicz}, {Fouqu{\'e}},
  {Hill}, {Kains}, {Kane}, {Marquette}, {Martin}, {Pollard}, {Sahu}, {Vinter},
  {Warren}, {Watson}, {Zub}, {Sumi}, {Szyma{\'n}ski}, {Kubiak}, {Poleski},
  {Soszynski}, {Ulaczyk}, {Pietrzy{\'n}ski}, \& {Wyrzykowski}}]{cassan}
{Cassan}, A., {Kubas}, D., {Beaulieu}, J.-P., {et~al.} 2012, \nat, 481, 167

\bibitem[{{Clanton} \& {Gaudi}(2014)}]{clanton2014}
{Clanton}, C., \& {Gaudi}, B.~S. 2014, \apj, 791, 90

\bibitem[{{Covone} {et~al.}(2000){Covone}, {de Ritis}, {Dominik}, \&
  {Marino}}]{covono2000}
{Covone}, G., {de Ritis}, R., {Dominik}, M., \& {Marino}, A.~A. 2000, \aap,
  357, 816

\bibitem[{{de Jong} {et~al.}(2006){de Jong}, {Widrow}, {Cseresnjes}, {Kuijken},
  {Crotts}, {Bergier}, {Baltz}, {Gyuk}, {Sackett}, {Uglesich}, \&
  {Sutherland}}]{dj2006}
{de Jong}, J.~T.~A., {Widrow}, L.~M., {Cseresnjes}, P., {et~al.} 2006, \aap,
  446, 855

\bibitem[{{Deb}(2017)}]{deb2017}
{Deb}, S. 2017, ArXiv e-prints, arXiv:1707.03130

\bibitem[{Dempster {et~al.}(1977)Dempster, Laird, \& Rubin}]{Dempster77}
Dempster, A.~P., Laird, N.~M., \& Rubin, D.~B. 1977, Journal of the Royal
  Statistical Society, Series B, 39, 1

\bibitem[{{Dong} {et~al.}(2007){Dong}, {Udalski}, {Gould}, {Reach}, {Christie},
  {Boden}, {Bennett}, {Fazio}, {Griest}, {Szyma{\'n}ski}, {Kubiak},
  {Soszy{\'n}ski}, {Pietrzy{\'n}ski}, {Szewczyk}, {Wyrzykowski}, {Ulaczyk},
  {Wieckowski}, {Paczy{\'n}ski}, {DePoy}, {Pogge}, {Preston}, {Thompson}, \&
  {Patten}}]{dong2007}
{Dong}, S., {Udalski}, A., {Gould}, A., {et~al.} 2007, \apj, 664, 862

\bibitem[{{Evans} \& {Howarth}(2008)}]{evans2008}
{Evans}, C.~J., \& {Howarth}, I.~D. 2008, \mnras, 386, 826

\bibitem[{{Gould}(2000)}]{gould2000}
{Gould}, A. 2000, \apj, 535, 928

\bibitem[{{Gould} \& {Loeb}(1992)}]{gould1992}
{Gould}, A., \& {Loeb}, A. 1992, \apj, 396, 104

\bibitem[{{Graff} \& {Gardiner}(1999)}]{graff1999}
{Graff}, D.~S., \& {Gardiner}, L.~T. 1999, \mnras, 307, 577

\bibitem[{{Griest} \& {Safizadeh}(1998)}]{griest1998}
{Griest}, K., \& {Safizadeh}, N. 1998, \apj, 500, 37

\bibitem[{{Harris} \& {Zaritsky}(2006)}]{harris2006}
{Harris}, J., \& {Zaritsky}, D. 2006, \aj, 131, 2514

\bibitem[{{Haschke} {et~al.}(2011){Haschke}, {Grebel}, \&
  {Duffau}}]{haschke2011}
{Haschke}, R., {Grebel}, E.~K., \& {Duffau}, S. 2011, \aj, 141, 158

\bibitem[{{Howard} {et~al.}(2012){Howard}, {Marcy}, {Bryson}, {Jenkins},
  {Rowe}, {Batalha}, {Borucki}, {Koch}, {Dunham}, {Gautier}, {Van Cleve},
  {Cochran}, {Latham}, {Lissauer}, {Torres}, {Brown}, {Gilliland}, {Buchhave},
  {Caldwell}, {Christensen-Dalsgaard}, {Ciardi}, {Fressin}, {Haas}, {Howell},
  {Kjeldsen}, {Seager}, {Rogers}, {Sasselov}, {Steffen}, {Basri},
  {Charbonneau}, {Christiansen}, {Clarke}, {Dupree}, {Fabrycky}, {Fischer},
  {Ford}, {Fortney}, {Tarter}, {Girouard}, {Holman}, {Johnson}, {Klaus},
  {Machalek}, {Moorhead}, {Morehead}, {Ragozzine}, {Tenenbaum}, {Twicken},
  {Quinn}, {Isaacson}, {Shporer}, {Lucas}, {Walkowicz}, {Welsh}, {Boss},
  {Devore}, {Gould}, {Smith}, {Morris}, {Prsa}, {Morton}, {Still}, {Thompson},
  {Mullally}, {Endl}, \& {MacQueen}}]{howard2012}
{Howard}, A.~W., {Marcy}, G.~W., {Bryson}, S.~T., {et~al.} 2012, \apjs, 201, 15

\bibitem[{{Ingrosso} {et~al.}(2009){Ingrosso}, {Novati}, {de Paolis}, {Jetzer},
  {Nucita}, \& {Zakharov}}]{ingrosso}
{Ingrosso}, G., {Novati}, S.~C., {de Paolis}, F., {et~al.} 2009, \mnras, 399,
  219

\bibitem[{{Jacklin} {et~al.}(2015){Jacklin}, {Lund}, {Pepper}, \&
  {Stassun}}]{lund2}
{Jacklin}, S., {Lund}, M.~B., {Pepper}, J., \& {Stassun}, K.~G. 2015, \aj, 150,
  34

\bibitem[{{Jacyszyn-Dobrzeniecka} {et~al.}(2016){Jacyszyn-Dobrzeniecka},
  {Skowron}, {Mr{\'o}z}, {Skowron}, {Soszy{\'n}ski}, {Udalski}, {Pietrukowicz},
  {Koz{\l}owski}, {Wyrzykowski}, {Poleski}, {Pawlak}, {Szyma{\'n}ski}, \&
  {Ulaczyk}}]{ania_cep}
{Jacyszyn-Dobrzeniecka}, A.~M., {Skowron}, D.~M., {Mr{\'o}z}, P., {et~al.}
  2016, \actaa, 66, 149

\bibitem[{{Jacyszyn-Dobrzeniecka} {et~al.}(2017){Jacyszyn-Dobrzeniecka},
  {Skowron}, {Mr{\'o}z}, {Soszy{\'n}ski}, {Udalski}, {Pietrukowicz}, {Skowron},
  {Poleski}, {Koz{\l}owski}, {Wyrzykowski}, {Pawlak}, {Szyma{\'n}ski}, \&
  {Ulaczyk}}]{ania_RR}
---. 2017, \actaa, 67, 1

\bibitem[{{Kallivayalil} {et~al.}(2013){Kallivayalil}, {van der Marel},
  {Besla}, {Anderson}, \& {Alcock}}]{kalli2013}
{Kallivayalil}, N., {van der Marel}, R.~P., {Besla}, G., {Anderson}, J., \&
  {Alcock}, C. 2013, \apj, 764, 161

\bibitem[{{Kiraga} \& {Paczynski}(1994)}]{kiraga1994}
{Kiraga}, M., \& {Paczynski}, B. 1994, \apjl, 430, L101

\bibitem[{{Kroupa}(2001)}]{kroupa2001}
{Kroupa}, P. 2001, \mnras, 322, 231

\bibitem[{{LSST Science Collaboration}(2009)}]{LSST}
{LSST Science Collaboration}. 2009, ArXiv e-prints, arXiv:0912.0201

\bibitem[{{Lund} {et~al.}(2015){Lund}, {Pepper}, \& {Stassun}}]{lund1}
{Lund}, M.~B., {Pepper}, J., \& {Stassun}, K.~G. 2015, \aj, 149, 16

\bibitem[{{Mao} \& {Paczynski}(1991)}]{mao1991}
{Mao}, S., \& {Paczynski}, B. 1991, \apjl, 374, L37

\bibitem[{{Marigo} {et~al.}(2017){Marigo}, {Girardi}, {Bressan}, {Rosenfield},
  {Aringer}, {Chen}, {Dussin}, {Nanni}, {Pastorelli}, {Rodrigues}, {Trabucchi},
  {Bladh}, {Dalcanton}, {Groenewegen}, {Montalb{\'a}n}, \& {Wood}}]{marigo2017}
{Marigo}, P., {Girardi}, L., {Bressan}, A., {et~al.} 2017, \apj, 835, 77

\bibitem[{{Mayor} \& {Queloz}(1995)}]{mayor}
{Mayor}, M., \& {Queloz}, D. 1995, \nat, 378, 355

\bibitem[{{Mayor} {et~al.}(2011){Mayor}, {Marmier}, {Lovis}, {Udry},
  {S{\'e}gransan}, {Pepe}, {Benz}, {Bertaux}, {Bouchy}, {Dumusque}, {Lo Curto},
  {Mordasini}, {Queloz}, \& {Santos}}]{mayor2011}
{Mayor}, M., {Marmier}, M., {Lovis}, C., {et~al.} 2011, ArXiv e-prints,
  arXiv:1109.2497

\bibitem[{{Mr{\'o}z} {et~al.}(2017){Mr{\'o}z}, {Udalski}, {Skowron}, {Poleski},
  {Koz{\l}owski}, {Szyma{\'n}ski}, {Soszy{\'n}ski}, {Wyrzykowski},
  {Pietrukowicz}, {Ulaczyk}, {Skowron}, \& {Pawlak}}]{mroz2017}
{Mr{\'o}z}, P., {Udalski}, A., {Skowron}, J., {et~al.} 2017, \nat, 548, 183

\bibitem[{{Muraveva} {et~al.}(2018){Muraveva}, {Subramanian}, {Clementini},
  {Cioni}, {Palmer}, {van Loon}, {Moretti}, {de Grijs}, {Molinaro}, {Ripepi},
  {Marconi}, {Emerson}, \& {Ivanov}}]{muraveva2017}
{Muraveva}, T., {Subramanian}, S., {Clementini}, G., {et~al.} 2018, \mnras,
  473, 3131

\bibitem[{{Paczynski}(1986)}]{paczynski}
{Paczynski}, B. 1986, \apj, 304, 1

\bibitem[{{Palanque-Delabrouille} {et~al.}(1998){Palanque-Delabrouille},
  {Afonso}, {Albert}, {Andersen}, {Ansari}, {Aubourg}, {Bareyre}, {Bauer},
  {Beaulieu}, {Bouquet}, {Char}, {Charlot}, {Couchot}, {Coutures}, {Derue},
  {Ferlet}, {Glicenstein}, {Goldman}, {Gould}, {Graff}, {Gros}, {Haissinski},
  {Hamilton}, {Hardin}, {de Kat}, {Lesquoy}, {Loup}, {Magneville}, {Mansoux},
  {Marquette}, {Maurice}, {Milsztajn}, {Moniez}, {Perdereau}, {Prevot},
  {Renault}, {Rich}, {Spiro}, {Vidal-Madjar}, {Vigroux}, {Zylberajch}, \& {EROS
  Collaboration}}]{palanque1998}
{Palanque-Delabrouille}, N., {Afonso}, C., {Albert}, J.~N., {et~al.} 1998,
  \aap, 332, 1

\bibitem[{{Penny} {et~al.}(2016){Penny}, {Henderson}, \& {Clanton}}]{penny2016}
{Penny}, M.~T., {Henderson}, C.~B., \& {Clanton}, C. 2016, \apj, 830, 150

\bibitem[{{Ripepi} {et~al.}(2017){Ripepi}, {Cioni}, {Moretti}, {Marconi},
  {Bekki}, {Clementini}, {de Grijs}, {Emerson}, {Groenewegen}, {Ivanov},
  {Molinaro}, {Muraveva}, {Oliveira}, {Piatti}, {Subramanian}, \& {van
  Loon}}]{ripepi2017}
{Ripepi}, V., {Cioni}, M.-R.~L., {Moretti}, M.~I., {et~al.} 2017, \mnras, 472,
  808

\bibitem[{{Sahu}(1994)}]{sahu1994}
{Sahu}, K.~C. 1994, \nat, 370, 275

\bibitem[{{Sahu} \& {Sahu}(1998)}]{sahu1998}
{Sahu}, K.~C., \& {Sahu}, M.~S. 1998, \apjl, 508, L147

\bibitem[{{Scowcroft} {et~al.}(2016){Scowcroft}, {Freedman}, {Madore},
  {Monson}, {Persson}, {Rich}, {Seibert}, \& {Rigby}}]{scowcroft2016}
{Scowcroft}, V., {Freedman}, W.~L., {Madore}, B.~F., {et~al.} 2016, \apj, 816,
  49

\bibitem[{{Shvartzvald} {et~al.}(2016){Shvartzvald}, {Maoz}, {Udalski}, {Sumi},
  {Friedmann}, {Kaspi}, {Poleski}, {Szyma{\'n}ski}, {Skowron}, {Koz{\l}owski},
  {Wyrzykowski}, {Mr{\'o}z}, {Pietrukowicz}, {Pietrzy{\'n}ski},
  {Soszy{\'n}ski}, {Ulaczyk}, {Abe}, {Barry}, {Bennett}, {Bhattacharya},
  {Bond}, {Freeman}, {Inayama}, {Itow}, {Koshimoto}, {Ling}, {Masuda}, {Fukui},
  {Matsubara}, {Muraki}, {Ohnishi}, {Rattenbury}, {Saito}, {Sullivan},
  {Suzuki}, {Tristram}, {Wakiyama}, \& {Yonehara}}]{yossi2016}
{Shvartzvald}, Y., {Maoz}, D., {Udalski}, A., {et~al.} 2016, \mnras, 457, 4089

\bibitem[{{Soszy{\'n}ski} {et~al.}(2015){Soszy{\'n}ski}, {Udalski},
  {Szyma{\'n}ski}, {Skowron}, {Pietrzy{\'n}ski}, {Poleski}, {Pietrukowicz},
  {Skowron}, {Mr{\'o}z}, {Koz{\l}owski}, {Wyrzykowski}, {Ulaczyk}, \&
  {Pawlak}}]{soszyn2015}
{Soszy{\'n}ski}, I., {Udalski}, A., {Szyma{\'n}ski}, M.~K., {et~al.} 2015,
  \actaa, 65, 297

\bibitem[{{Soszy{\'n}ski} {et~al.}(2016){Soszy{\'n}ski}, {Udalski},
  {Szyma{\'n}ski}, {Wyrzykowski}, {Ulaczyk}, {Poleski}, {Pietrukowicz},
  {Koz{\l}owski}, {Skowron}, {Skowron}, {Mr{\'o}z}, \& {Pawlak}}]{soszyn2016}
---. 2016, \actaa, 66, 131

\bibitem[{{Stanimirovi{\'c}} {et~al.}(2004){Stanimirovi{\'c}},
  {Staveley-Smith}, \& {Jones}}]{stani2004}
{Stanimirovi{\'c}}, S., {Staveley-Smith}, L., \& {Jones}, P.~A. 2004, \apj,
  604, 176

\bibitem[{{Szkody} {et~al.}(2011){Szkody}, {Long}, {DiStefano}, {Henden},
  {Kalirai}, {Kashyap}, {Kasliwal}, {Smith}, \& {Stassun}}]{szkody}
{Szkody}, P., {Long}, K.~S., {DiStefano}, R., {et~al.} 2011, {Science White
  Paper for LSST Deep-Drilling Field Observations High Cadence Observations of
  the Magellanic Clouds and Select Galactic Cluster Fields} (Tucson, AZ : LSST)

\bibitem[{{Tisserand} {et~al.}(2007){Tisserand}, {Le Guillou}, {Afonso},
  {Albert}, {Andersen}, {Ansari}, {Aubourg}, {Bareyre}, {Beaulieu}, {Charlot},
  {Coutures}, {Ferlet}, {Fouqu{\'e}}, {Glicenstein}, {Goldman}, {Gould},
  {Graff}, {Gros}, {Haissinski}, {Hamadache}, {de Kat}, {Lasserre}, {Lesquoy},
  {Loup}, {Magneville}, {Marquette}, {Maurice}, {Maury}, {Milsztajn}, {Moniez},
  {Palanque-Delabrouille}, {Perdereau}, {Rahal}, {Rich}, {Spiro},
  {Vidal-Madjar}, {Vigroux}, {Zylberajch}, \& {EROS-2
  Collaboration}}]{tisserand2007}
{Tisserand}, P., {Le Guillou}, L., {Afonso}, C., {et~al.} 2007, \aap, 469, 387

\bibitem[{{Udalski} {et~al.}(1993){Udalski}, {Szymanski}, {Kaluzny}, {Kubiak},
  {Krzeminski}, {Mateo}, {Preston}, \& {Paczynski}}]{udalski1993}
{Udalski}, A., {Szymanski}, M., {Kaluzny}, J., {et~al.} 1993, \actaa, 43, 289

\bibitem[{{Udalski} {et~al.}(1994){Udalski}, {Szymanski}, {Kaluzny}, {Kubiak},
  {Mateo}, {Krzeminski}, \& {Paczynski}}]{udalskiews}
---. 1994, \actaa, 44, 227

\bibitem[{{Udalski} {et~al.}(2015){Udalski}, {Szyma{\'n}ski}, \&
  {Szyma{\'n}ski}}]{udalski2015}
{Udalski}, A., {Szyma{\'n}ski}, M.~K., \& {Szyma{\'n}ski}, G. 2015, \actaa, 65,
  1

\bibitem[{{van der Marel} {et~al.}(2009){van der Marel}, {Kallivayalil}, \&
  {Besla}}]{vdm2009}
{van der Marel}, R.~P., {Kallivayalil}, N., \& {Besla}, G. 2009, in IAU
  Symposium, Vol. 256, The Magellanic System: Stars, Gas, and Galaxies, ed.
  J.~T. {Van Loon} \& J.~M. {Oliveira}, 81--92

\bibitem[{{van der Marel} \& {Sahlmann}(2016)}]{vdm2016}
{van der Marel}, R.~P., \& {Sahlmann}, J. 2016, \apjl, 832, L23

\bibitem[{{Wolszczan} \& {Frail}(1992)}]{wolszczan}
{Wolszczan}, A., \& {Frail}, D.~A. 1992, \nat, 355, 145

\bibitem[{{Wyrzykowski} {et~al.}(2009){Wyrzykowski}, {Koz{\l}owski}, {Skowron},
  {Belokurov}, {Smith}, {Udalski}, {Szyma{\'n}ski}, {Kubiak},
  {Pietrzy{\'n}ski}, {Soszy{\'n}ski}, {Szewczyk}, \&
  {{\.Z}ebru{\'n}}}]{wyrzyk2009}
{Wyrzykowski}, {\L}., {Koz{\l}owski}, S., {Skowron}, J., {et~al.} 2009, \mnras,
  397, 1228

\bibitem[{{Wyrzykowski} {et~al.}(2010){Wyrzykowski}, {Koz{\l}owski}, {Skowron},
  {Belokurov}, {Smith}, {Udalski}, {Szyma{\'n}ski}, {Kubiak},
  {Pietrzy{\'n}ski}, {Soszy{\'n}ski}, \& {Szewczyk}}]{wyrzyk2010}
---. 2010, \mnras, 407, 189

\bibitem[{{Wyrzykowski} {et~al.}(2011{\natexlab{a}}){Wyrzykowski},
  {Koz{\l}owski}, {Skowron}, {Udalski}, {Szyma{\'n}ski}, {Kubiak},
  {Pietrzy{\'n}ski}, {Soszy{\'n}ski}, {Szewczyk}, {Ulaczyk}, \&
  {Poleski}}]{wyrzyk2011_L}
---. 2011{\natexlab{a}}, \mnras, 413, 493

\bibitem[{{Wyrzykowski} {et~al.}(2011{\natexlab{b}}){Wyrzykowski}, {Skowron},
  {Koz{\l}owski}, {Udalski}, {Szyma{\'n}ski}, {Kubiak}, {Pietrzy{\'n}ski},
  {Soszy{\'n}ski}, {Szewczyk}, {Ulaczyk}, {Poleski}, \&
  {Tisserand}}]{wyrzyk2011}
{Wyrzykowski}, {\L}., {Skowron}, J., {Koz{\l}owski}, S., {et~al.}
  2011{\natexlab{b}}, \mnras, 416, 2949

\end{thebibliography}

\end{document}